\documentclass[prx, aps, showpacs, twocolumn, preprintnumbers, amsmath, amssymb, superscriptaddress, nofootinbib, longbibliography, nofootinbibjustification=justified, singlelinecheck=false, 10pt]{revtex4-2}

\usepackage{amsfonts}
\usepackage{graphicx}
\usepackage{hyperref} 
\usepackage{cleveref}
\usepackage{enumerate}
\usepackage{bm}

\usepackage{tikz}
\usepackage{circuitikz}
\usetikzlibrary{quantikz2,shapes.geometric}
\definecolor{myorange}{HTML}{FF8001}
\definecolor{myblue}{HTML}{99C1F1}
\tikzset{
  mysquare/.style={draw, fill=myblue, minimum size=3.2mm, inner sep=0pt},
  mydiamond/.style={draw, fill=myorange, diamond, aspect=1, minimum size=4mm, inner sep=0pt},
  mycircle/.style={draw, fill=gray!40, circle, minimum size=3.6mm, inner sep=0pt},
  cubedetector/.style={
    shape=rectangle,
    draw,
    fill=black!90,
    minimum width=7mm,      % 2 units in x-direction
    minimum height=7mm,     % 1 unit in y-direction
    inner sep=0pt
  },
  flatdetector/.style={
    shape=rectangle,
    draw,
    fill=black!90,
    minimum width=7mm,      % 2 units in x-direction
    minimum height=5mm,     % 1 unit in y-direction
    inner sep=0pt
  }
}

\newcommand{\sortednoon}{
\begin{tikzpicture}[x=0.8cm,y=-0.7cm] % tighter vertical spacing
    \draw (0,0) -- (7,0); % wires
    \draw (0,0.8) -- (7,0.8); % wires
    \node[left] at (0,0) {$\ket{0}$}; % initial vacuum state
    \node[left] at (0,0.8) {$\ket{0}$}; % initial vacuum state
    \node[cubedetector] at (7.4,0.4) {\textcolor{white}{\bm{$=$}\textbf{N}}}; % final detector
    \draw (0,2) -- (7,2); % wires
    \node[left] at (0,2) {$\ket{0}$}; % initial vacuum state
    \node[flatdetector] at (7.4,2) {\textcolor{white}{\bm{$\geq \tau$}}}; % final detector

    \pgfmathsetmacro{\xnow}{.9}

    \node[mycircle, very thick] at (\xnow,0) {};
    \node[above=4mm] at (\xnow,0) {A};
    \pgfmathsetmacro{\xnow}{\xnow + 1}
    
    \draw[very thick] (\xnow,0) -- (\xnow,0.8); % Vertical connection first
    \node[mycircle, very thick] at (\xnow,0) {};  % Gates on top
    \node[mycircle, very thick] at (\xnow,0.8) {};
    \node[above=4mm] at (\xnow,0) {B};
    \pgfmathsetmacro{\xnow}{\xnow + 1}
    
    \draw[very thick] (\xnow,0) -- (\xnow,2); % Vertical connection first
    \node[mycircle, very thick] at (\xnow,0) {};  % Gates on top
    \node[mycircle, very thick] at (\xnow,2) {};
    \node[above=4mm] at (\xnow,0) {C};
    \pgfmathsetmacro{\xnow}{\xnow + 1}

    \node[mycircle, very thick] at (\xnow,0.8) {};
    \node[above=4mm] at (\xnow,0) {D};
    \pgfmathsetmacro{\xnow}{\xnow + 1}

    \draw[very thick] (\xnow,0.8) -- (\xnow,2); % Vertical connection first
    \node[mycircle, very thick] at (\xnow,0.8) {};  % Gates on top
    \node[mycircle, very thick] at (\xnow,2) {};
    \node[above=4mm] at (\xnow,0) {E};
    \pgfmathsetmacro{\xnow}{\xnow + 1}

    \node[mycircle, very thick] at (\xnow,2) {};
    \node[above=4mm] at (\xnow,0) {F};
    \pgfmathsetmacro{\xnow}{\xnow + 1}
\end{tikzpicture}
}

\newcommand{\originalwstate}{ % ((0, 7), (0, 5), (0, 3), (2, 4), (2, 6), (1, 2), (4, 6))
\begin{tikzpicture}[x=0.8cm,y=-0.8cm] % tighter vertical spacing
  \foreach \y in {0,...,3} {
    \draw (0,\y) -- (7.8,\y); % wires
    \node[left] at (0,\y) {$\ket{0,\!0}$}; % initial vacuum state
    \node[flatdetector] at (7.5,\y) {\textcolor{white}{\bm{$\geq1$}}}; % final detector
  }
  \pgfmathsetmacro{\xnow}{0.8}
  
  \draw[very thick] (\xnow,0) -- (\xnow,2); % Vertical connection first
  \node[mysquare, very thick] at (\xnow,0) {};  % Gates on top
  \node[mydiamond, very thick] at (\xnow,2) {};
  \node[above=4mm] at (\xnow,0) {A}; % Text above gates
  \pgfmathsetmacro{\xnow}{\xnow + 0.9}

\draw[very thick] (\xnow,0) -- (\xnow,3); % Vertical connection first
  \node[mysquare, very thick] at (\xnow,0) {};  % Gates on top
  \node[mydiamond, very thick] at (\xnow,3) {};
  \node[above=4mm] at (\xnow,0) {B}; % Text above gates
  \pgfmathsetmacro{\xnow}{\xnow + 0.9}
  
  \draw[very thick] (\xnow,0) -- (\xnow,1); % Vertical connection first
  \node[mysquare, very thick] at (\xnow,0) {};  % Gates on top
  \node[mydiamond, very thick] at (\xnow,1) {};
  \node[above=4mm] at (\xnow,0) {C}; % Text above gates
  \pgfmathsetmacro{\xnow}{\xnow + 0.9}
  
  \draw[very thick] (\xnow,1) -- (\xnow,3); % Vertical connection first
  \node[mysquare, very thick] at (\xnow,1) {};  % Gates on top
  \node[mysquare, very thick] at (\xnow,3) {};
  \node[above=4mm] at (\xnow,0) {D}; % Text above gates
  \pgfmathsetmacro{\xnow}{\xnow + 0.9}

  \draw[very thick] (\xnow,1) -- (\xnow,2); % Vertical connection first
  \node[mysquare, very thick] at (\xnow,1) {};  % Gates on top
  \node[mysquare, very thick] at (\xnow,2) {}; 
  \node[above=4mm] at (\xnow,0) {E}; % Text above gates
  \pgfmathsetmacro{\xnow}{\xnow + 0.9}

  \draw[very thick] (\xnow,0) -- (\xnow,1); % Vertical connection first
  \node[mydiamond, very thick] at (\xnow,0) {};  % Gates on top
  \node[mysquare, very thick] at (\xnow,1) {};
  \node[above=4mm] at (\xnow,0) {F}; % Text above gates
  \pgfmathsetmacro{\xnow}{\xnow + 0.9}
  
  \draw[very thick] (\xnow,2) -- (\xnow,3); % Vertical connection first
  \node[mysquare, very thick] at (\xnow,2) {};  % Gates on top
  \node[mysquare, very thick] at (\xnow,3) {};
  \node[above=4mm] at (\xnow,0) {G}; % Text above gates
  \pgfmathsetmacro{\xnow}{\xnow + 0.9}

\end{tikzpicture}
}

\tikzset{
    detector/.pic={
        \draw[fill] (0,0) circle [radius=5pt];% circle
        \draw[fill] (-7pt,-5pt) rectangle (-1pt,5pt); % rectangle (centered)
    }
}
\tikzset{
  modulebox/.pic={
    \tikzset{modulebox/.cd, #1} % load arguments into keys
    \begin{scope}[]
      \draw[
        very thick,
        fill=\moduleboxfill,
        rounded corners=\moduleboxradius
      ]
        (-.6,-0.75) rectangle (.6,0.75);
      \node at (0,0) {\moduleboxtext};
    \end{scope}
  },
  modulebox/.cd,
    text/.store in=\moduleboxtext,
    fill/.store in=\moduleboxfill,
    radius/.store in=\moduleboxradius,
    text={},              % defaults
    fill=gray!20,
    radius=0pt
}
\newcommand{\BLnoonFourModifiedGraph}{
\begin{tikzpicture}[
    every path/.style={line width=1pt},
    myloop/.style={
            loop,           % use TikZ’s native loop shape
            min distance=10mm,  % default loop size
            -,              % no arrow
            edgelabel/.style={inner sep=4pt}%, font=\scriptsize} % label style
        },
    graphnode/.style={
        circle, draw,
        inner sep=1.5pt,
    },
    ancillary/.style={
        rectangle, draw,
        minimum width=4mm,
        minimum height=4mm,
        inner sep=1.5pt,
    },
    edgelabel/.style={inner sep=4pt},%,line width=1pt},
]
\node[graphnode] (A) at (0,0) {0};
\node[graphnode] (B) at (2,0) {1};
\node[ancillary] (C) at (1,-1.7) {2};

\draw (B) -- node[edgelabel, right] {$e=1$} (C);

\draw (A) edge[myloop, loop, in=180, out=120, looseness=6] node[edgelabel,left] {$a=2$} (A);
\draw (B) edge[myloop, loop, in=15, out=75, looseness=6] node[edgelabel,right] {$d=2$} (B);
\draw (C) edge[myloop, loop, in=195, out=255, looseness=6] node[edgelabel,left] {$f=-1$} (C);
\end{tikzpicture}
}

\newcommand{\BLnoonFourGraph}{
\begin{tikzpicture}[
    every path/.style={line width=1pt},
    myloop/.style={
            loop,           % use TikZ’s native loop shape
            min distance=10mm,  % default loop size
            -,              % no arrow
            edgelabel/.style={inner sep=4pt}%, font=\scriptsize} % label style
        },
    graphnode/.style={
        circle, draw,
        inner sep=1.5pt,
    },
    ancillary/.style={
        rectangle, draw,
        minimum width=4mm,
        minimum height=4mm,
        inner sep=1.5pt,
    },
    edgelabel/.style={inner sep=4pt},
]
\node[graphnode] (0) at (0,0) {0};
\node[graphnode] (1) at (2,0) {1};
\node[ancillary] (2) at (0,-2) {2};
\node[ancillary] (3) at (2,-2) {3};

\draw (1) -- node[edgelabel, above] {} (2);
\draw (2) -- node[edgelabel, left]  {} (3);
\draw (1) -- node[edgelabel, right] {} (3);

\draw (A) edge[myloop, loop, in=180, out=120, looseness=6] node[edgelabel,left] {} (A);
\draw (B) edge[myloop, loop, in=0, out=60, looseness=6] node[edgelabel,right] {} (B);
\end{tikzpicture}}

\newcommand{\BLnoonThreeGraph}{
\begin{tikzpicture}[
    every path/.style={line width=1pt},
    myloop/.style={
            loop,           % use TikZ’s native loop shape
            min distance=10mm,  % default loop size
            -,              % no arrow
            edgelabel/.style={inner sep=4pt}%, font=\scriptsize} % label style
        },
    graphnode/.style={
        circle, draw,
        inner sep=1.5pt,
    },
    ancillary/.style={
        rectangle, draw,
        minimum width=4mm,
        minimum height=4mm,
        inner sep=1.5pt,
    },
    edgelabel/.style={inner sep=4pt,line width=1pt},
]
\node[graphnode] (A) at (0,0) {0};
\node[graphnode] (B) at (2,0) {1};
\node[ancillary] (C) at (1,-1.7) {2};

\draw (A) -- node[edgelabel, above] {$b=-1$} (B);
\draw (A) -- node[edgelabel, left]  {$c=1$} (C);
\draw (B) -- node[edgelabel, right] {$e=1$} (C);

\draw (A) edge[myloop, loop, in=180, out=120, looseness=6] node[edgelabel,left] {$a=2$} (A);
\draw (B) edge[myloop, loop, in=0, out=60, looseness=6] node[edgelabel,right] {$d=2$} (B);
\end{tikzpicture}}

\newcommand{\HeraldedBellCircuitFour}{
\begin{circuitikz}[scale=0.8, transform shape]
\tikzset{
  every node/.style={font=\large},
  every path/.style={line width=1pt}
}
\draw [ line width=1.75pt , rounded corners = 7.8, dashed] (-3.35,1.15) rectangle  (-1.8,-0.9);
\draw (-4,1.75) -- (-4,1.75);
\draw (-3.5,2) -- (-1.75,2);
\draw (-3.5,0.75) -- (-1.75,0.75);
\draw (-0.5,2) -- (0.75,2);
\draw (-0.5,0.75) -- (3.25,0.75);
\draw (-0.5,-0.5) -- (3.25,-0.5);
\draw (-0.5,-1.75) -- (0.75,-1.75);
\draw [line width=1.5pt] (1.5,1.25) rectangle (2.5,0.25);
\draw [line width=1.5pt] (1.5,0) rectangle (2.5,-1);
\draw [line width=1.5pt] (1.5,0) -- (2.5,-1);
\draw [line width=1.5pt] (2.5,1.25) -- (1.5,0.25);
\draw [line width=1.5pt] (0.25,0) rectangle (0.5,-1);
\draw (2,-0.5) -- (2,-1.75);
\draw (2,2) -- (2,0.75);
\draw (-3.5,-0.5) -- (-1.75,-0.5);
\draw (-3.5,-1.75) -- (-1.75,-1.75);
\draw (-1.75,-1.75) .. controls (-1,-1.75) and (-1.25,-0.5) .. (-0.5,-0.5);
\draw (-0.5,-1.75) .. controls (-1.25,-1.75) and (-1,-0.5) .. (-1.75,-0.5);
\draw (-1.75,0.75) .. controls (-1,0.75) and (-1.25,2) .. (-0.5,2);
\draw (-0.5,0.75) .. controls (-1.25,0.75) and (-1,2) .. (-1.75,2);
\draw [line width=2pt](-1.6,-1.13) -- (-0.6,-1.13);
\draw [line width=2pt](-1.6,1.4) -- (-0.6,1.4);

\node at (-2.6,1.5) {SPDC};
\node at (-0.1,1.55) {BS1};
\node at (0.8,2.3) {\textit{\textbf{out}}};
\node at (0.35,0.3) {HWP};
\node at (-0.1,-1.3) {BS2};
\node at (0.8,-1.95) {\textit{\textbf{out}}};
\node at (2.75,-1.35) {PBS2};
\node at (2.75,1.6) {PBS1};

\node at (-4,2) {$\ket{0,\!0}$};
\node at (-4,.75) {$\ket{0,\!0}$};
\node at (-4,-.5) {$\ket{0,\!0}$};
\node at (-4,-1.75) {$\ket{0,\!0}$};

\pic[scale=1.25] at (3.5,0.75) {detector};
\pic[scale=1.25] at (3.5,-0.5) {detector};
\pic[scale=1.25,rotate=90] at (2,2.25) {detector};
\pic[scale=1.25,rotate=-90] at (2,-2) {detector};

\draw[line width=1.5pt](-2.95,0.75) -- (-2.95,-0.5);
\draw[line width=1.5pt](-2.2,0.75) -- (-2.2,-0.5);
  \node[mysquare, very thick, scale=1.25] at (-2.95,0.75) {};  % Gates on top
  \node[mydiamond, very thick, scale=1.25] at (-2.95,-0.5) {};
  \node[mydiamond, very thick, scale=1.25] at (-2.2,0.75) {};
  \node[mysquare, very thick, scale=1.25] at (-2.2,-0.5) {};  % Gates on top
\end{circuitikz}}

\newcommand{\HeraldedBellCircuitEight}{
\begin{circuitikz}[scale=0.8, transform shape]
\tikzset{
  every node/.style={font=\large},
  every path/.style={line width=1pt}
}
\draw (-3.25,4.5) -- (-1.5,4.5);
\draw (-3.25,2) -- (-1.5,2);
\draw (1,2) .. controls (0,2) and (-0.5,4.5) .. (-1.5,4.5);
\draw (-1.5,2) .. controls (-0.5,2) and (0,4.5) .. (1,4.5);
\draw (-3.25,3.25) -- (-1.5,3.25);
\draw (-3.25,0.75) -- (-1.5,0.75);
\draw (1,0.75) .. controls (0,0.75) and (-0.5,3.25) .. (-1.5,3.25);
\draw (-1.5,0.75) .. controls (-0.5,0.75) and (0,3.25) .. (1,3.25);
\draw (-3.25,-0.5) -- (-1.5,-0.5);
\draw (-3.25,-3) -- (-1.5,-3);
\draw (1,-3) .. controls (0,-3) and (-0.5,-0.5) .. (-1.5,-0.5);
\draw (-1.5,-3) .. controls (-0.5,-3) and (0,-0.5) .. (1,-0.5);
\draw (-3.25,-1.75) -- (-1.5,-1.75);
\draw (-3.25,-4.25) -- (-1.5,-4.25);
\draw (1,-4.25) .. controls (0,-4.25) and (-0.5,-1.75) .. (-1.5,-1.75);
\draw (-1.5,-4.25) .. controls (-0.5,-4.25) and (0,-1.75) .. (1,-1.75);
\draw (1,-0.5) -- (1.5,-0.5);
\draw (1,-1.75) -- (1.5,-1.75);
\draw (2.75,-0.5) -- (3.5,-0.5);
\draw (2.75,-1.75) -- (3.5,-1.75);
\draw (2.75,-1.75) .. controls (2,-1.75) and (2.25,-0.5) .. (1.5,-0.5);
\draw (1.5,-1.75) .. controls (2.25,-1.75) and (2,-0.5) .. (2.75,-0.5);
\draw (1,0.75) -- (3.5,0.75);
\draw (1,2) -- (3.5,2);
\draw (1,4.5) -- (3.33,4.5);
\draw (1,3.25) -- (3.33,3.25);
\draw (1,-3) -- (3.33,-3);
\draw (1,-4.25) -- (3.33,-4.25);
\node[flatdetector,scale=1.25, font=\normalsize] at (3.7,-.5) {\textcolor{white}{\bm{$\geq1$}}};
\node[flatdetector,scale=1.25, font=\normalsize] at (3.7,-1.75) {\textcolor{white}{\bm{$\geq1$}}};
\node[flatdetector,scale=1.25, font=\normalsize] at (3.7,.75) {\textcolor{white}{\bm{$\geq1$}}};
\node[flatdetector,scale=1.25, font=\normalsize] at (3.7,2) {\textcolor{white}{\bm{$\geq1$}}};

\draw[line width=1.4pt](-2.75,-1.75) -- (-2.75,2);
  \node[mycircle, very thick, scale=1.25] at (-2.75,-1.75) {};  % Gates on top
  \node[mycircle, very thick, scale=1.25] at (-2.75,2) {};

\draw[line width=1.4pt](-2.,0.75) -- (-2.,-0.5);
  \node[mycircle, very thick, scale=1.25] at (-2.,0.75) {};
  \node[mycircle, very thick, scale=1.25] at (-2.,-0.5) {};  % Gates on top

\draw [line width=1.75pt , rounded corners = 7.8, dashed] (-1.05,3.65) rectangle (.5,1.6);
\draw [line width=2pt](-0.75,3.25) -- (0.25,3.25);
\draw [line width=2pt](-0.75,2) -- (0.25,2);
\node at (-.25,4.1) {BS1};

\draw [line width=1.75pt , rounded corners = 7.8, dashed] (-1.05,-1.35) rectangle  (.5,-3.4);
\draw [line width=2pt](-0.75,-1.75) -- (0.25,-1.75);
\draw [line width=2pt](-0.75,-3) -- (0.25,-3);
\node at (-.25,-.9) {BS2};

\draw [line width=1.75pt, rounded corners = 7.8, dashed] (1.4,-0.3) rectangle  (2.85,-1.95);
\draw [line width=2pt](1.625,-1.125) -- (2.625,-1.125);
\node at (2.125,0.1) {HWP};

\draw [line width=1.75pt , rounded corners = 7.8, dashed] (-3.15,-2.15) rectangle  (-1.6,2.4);
\node at (-2.38,2.75) {SPDC};

\node at (-3.6,4.5) {$\ket{0}$};
\node at (-3.6,3.25) {$\ket{0}$};
\node at (-3.6,2) {$\ket{0}$};
\node at (-3.6,0.75) {$\ket{0}$};
\node at (-3.6,-0.5) {$\ket{0}$};
\node at (-3.6,-1.75) {$\ket{0}$};
\node at (-3.6,-3) {$\ket{0}$};
\node at (-3.6,-4.25) {$\ket{0}$};
\end{circuitikz}}

\usepackage{listings}
\usepackage{color}

\hypersetup{
    hidelinks,
    colorlinks=true,
    breaklinks=true,
    citecolor=SteelBlue,
    filecolor=LimeGreen,
    linkcolor=MediumBlue,
    urlcolor=MediumPurple
}
\usepackage[svgnames]{xcolor}

\newcommand{\mysection}[1]{%
  \phantomsection
  \pdfbookmark[1]{#1}{sec:\detokenize{#1}}%
  \textit{#1}---%
}

\begin{document}

\title{Automated experimental design for high-probability entanglement generation}

\author{Carlos Ruiz-Gonzalez}
\affiliation{Max Planck Institute for the Science of Light, Erlangen, Germany}
\affiliation{ELLIS Unit Linz, Institute for Machine Learning, JKU Linz, Austria}
\email{cruizgo@proton.me}

\author{Mario Krenn}
\affiliation{Max Planck Institute for the Science of Light, Erlangen, Germany}
\affiliation{Machine Learning in Science Cluster, Department for Computer Science, Faculty of Science, University of Tuebingen, Tuebingen, Germany}
\email{mario.krenn@uni-tuebingen.de}

\author{Xuemei Gu}
\affiliation{Institut für Festkörpertheorie und Optik, Friedrich-Schiller-Universität Jena, Jena, Germany}
\email{xuemei.gu@uni-jena.de}

\begin{abstract}
Entangled photons are widely used in quantum technologies. 
Many photonic experiments generate them with probabilistic photon-pair sources that can be modeled as squeeze operators. 
In practice, these sources are usually treated in the low-gain (perturbative) regime, keeping only the leading single-pair term and neglecting higher-order multi-pair emission events.
In pursuit of fidelity, the probability of successful entanglement generation can become extremely small, a tradeoff often ignored.
Here we develop an automated design algorithm for quantum experiments to optimize both fidelity and success probability while accounting for higher-order multi-pair emissions.
Our discovery algorithm explores different design topologies subject to varying hardware constraints.
It optimizes the source parameters to reduce undesired higher-order terms or even benefit from them.
The experiments presented outperform previous proposals for widely used states, including heralded Bell states, W states, and NOON states, paving the way for more efficient photonic technologies.
\end{abstract}
\maketitle
\mysection{Introduction}
Entanglement is a crucial resource in quantum technologies~\cite{horodecki2009entanglement}, enabling quantum computing~\cite{bourassa2021blueprint, xanadu2025scaling, psiquantum2025}, sensing~\cite{huang2024entanglement, ma2025unraveling}, and communications~\cite{azuma2023repeaters, zhang2025towards}.
Photonics is a promising platform for these advances~\cite{luo2023recent, wang2025scalable}: photons can travel long distances with low decoherence, often without cryogenic cooling~\cite{vanleent2022entangling, krvzivc2023towards}, and benefit from mature technologies such as optical fiber and integrated circuits~\cite{kucera2024demonstration, pittaluga2025long, nielsen2025programmable}.
To generate entangled photons, many setups combine probabilistic sources, such as spontaneous parametric down-conversion (SPDC), with linear elements and different detection schemes~\cite{ruiz2023digital, forbes2025heralded}.

When pumped with strong fields, SPDC sources can produce a high number of photon pairs~\cite{chizhov1993photon}.
Yet, most experiments operate in the perturbative low-gain regime.
The pump is so weak that multi-pair emission events are negligible, and each source is approximated by a single pair-creation term. Successful state generation is then identified by specific detector-click patterns~\cite{forbes2025heralded}. 
However, as the reported success probability is conditional on those detection outcomes, a fundamental tradeoff is overlooked.
As the pump decreases to preserve fidelity, the probability of obtaining the required detection pattern becomes extremely small.

Here, we model SPDC sources as full squeeze operators~\cite{quesada2022beyond, bulmer2022threshold, chinni2024beyond, gu2025analytical}, revealing that the photon statistics neglected by the perturbative approximation are central for efficient entanglement generation.
This turns experimental design into a joint optimization of two competing properties: fidelity and success probability.
To improve the tradeoff, we engineer interference between higher-order contributions, reducing undesired emissions. % that degrade the target state.
Interestingly, we see that higher-order events can enhance the experiments' performance when using ancillary photons.

We first revisit the generation of GHZ states to illustrate how the tradeoff emerges beyond the low-gain approximation.
We then present our algorithm to optimize the experiments' parameters and configurations, maximizing the success probability for a range of target fidelities. 
Finally, we apply the methods to produce a 4-qubit W state, a heralded Bell state, and $N=3$ and $N=4$ N00N states, surpassing previous proposals~\cite{graphs3, sliwa2003conditional, ruiz2023digital}.
These are widely used states with different entanglement structures, detection schemes, and photon statistics.
By moving beyond the low-gain approximation, our work sheds new light in probabilistic entanglement generation.

\begin{figure}[t]
    \includegraphics[width=\linewidth]{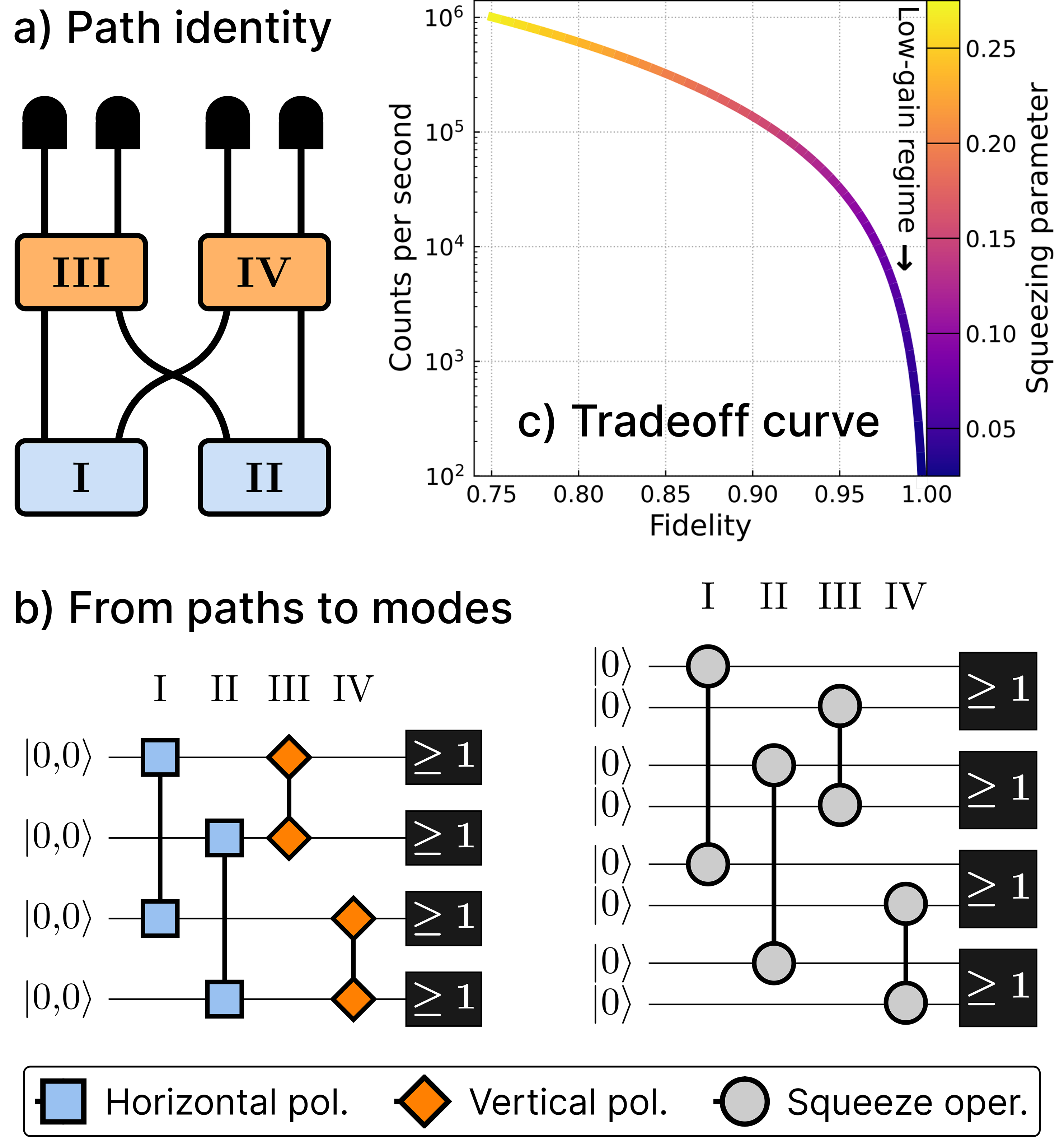}
    \caption{\textbf{Generating a 4-qubit GHZ state.} 
    \textbf{(a)} We replicate an entanglement generation experiment with SPDC crystals and threshold detectors~\cite{krenn2017pathidentity}.
    When all detectors click, the projected state is a superposition of horizontally and vertically polarized photons from the blue (I \& II) and orange (III \& IV) sources, respectively.
    With two Fock modes, we encode the photons as $\ket{\text{H}}\!=\!\ket{1,0}$ and $\ket{\text{V}}\!=\!\ket{0,1}$.
    \textbf{(b)} The SPDC crystals act as squeeze operators on 4 different optical paths, initialized at vacuum with 2 polarization modes: $\ket{0,0}$. 
    Splitting each path into 2 modes, we see the sources commute, as they act on different modes.
    \textbf{(c)} Increasing the common pump $r$, we transcend the low-gain approximation and reveal the tradeoff between fidelity and success probability.
    Here and in following figures, the shown counts per second assume the crystals are pumped by a laser with 10ns repetition rate.}
    \label{fig:ghz4circuits}
\end{figure}

\mysection{Beyond the low-gain regime}
The setup of Fig.~\ref{fig:ghz4circuits}, introduced in the low-gain regime~\cite{krenn2017pathidentity}, generates a 4-qubit GHZ state, $(\ket{\text{HHHH}} + \ket{\text{VVVV}})/\sqrt 2$, from four SPDC sources.
Since the photodetectors ignore polarization, we cannot know which sources fired when each detector clicks.
This indistinguishability of emission events is the basis of path identity experiments~\cite{zou91induced, wang91induced, kysela2020path, hochrainer2022quantum, barreto2022quantum, wang2024entangling, hu2025observation, bernecker2025quantum}.
In Fig.~\ref{fig:ghz4circuits}a, with a common pump $r\ll 1$ and phase $\theta=0$, each source can be approximated as $S_2(r,\theta)=\exp(r(e^{-i\theta}ab-e^{i\theta}a^\dagger b^\dagger))\approx(1+rab-ra^\dagger b^\dagger)$, where $a$ ($a^\dagger$) and $b$ ($b^\dagger$) are annihilation (creation) operators acting in two different modes.
The resulting state is, approximately,
\begin{align}
\label{eq:ghz4approx}
    \ket\psi=&S_\text{I}S_\text{II}S_\text{III}S_\text{IV}\ket{vac} \nonumber\\
    \approx&(1+rh_1h_3-rh_1^\dagger h_3^\dagger)(1+rh_2h_4-rh_2^\dagger h_4^\dagger) \nonumber\\
    &(1+rv_1v_2-rv_1^\dagger v_2^\dagger)(1+rv_3v_4-rv_3^\dagger v_4^\dagger)\ket{vac}, 
\end{align}
where $h_j^\dagger$ ($v_j^\dagger$) is the creation operator of horizontally (vertically) polarized photons in the $j-$th optical path.
When the four threshold detectors click, the final state is
\begin{align}
    |\tilde \psi\rangle =\:& r^2(h_1^\dagger h_2^\dagger h_3^\dagger h_4^\dagger + v_1^\dagger v_2^\dagger v_3^\dagger v_4^\dagger)\ket{vac}  \nonumber\\
    &-r^3(h_1^\dagger h_2^\dagger h_3^\dagger h_4^\dagger v_1^\dagger v_2^\dagger + \dots) \ket{vac}  + \dots\nonumber\\\approx\:&r^2(\ket{\text{HHHH}} + \ket{\text{VVVV}})
\label{eq:lowgainGHZ}.
\end{align}
Hence, in the low-gain approximation, terms with extra photons are neglected, annihilation operators can be omitted, and the fidelity is assumed optimal.

What the approximation obscures is how unlikely successful detection is, since the dominant terms with unpopulated paths are left out.
Beyond this low-gain regime, when the pump grows, so does the success probability.
At the same time, as the weight of high-order terms increases, the fidelity drops.
This tradeoff is shown in Fig.~\ref{fig:ghz4circuits}c, where the approximation from Eq.~\eqref{eq:lowgainGHZ} is only a small part of the picture.

To recompute this experiment and find our new proposals, we model the SPDC sources as full squeeze operators in the Fock basis. We compute a (truncated) infinite series of amplitudes for arbitrary emission and absorption events~\cite{dasgupta1996disentanglement, gu2025analytical}.
The two-mode SPDC is described as the following operator (detailed amplitudes in App.~\ref{app:squeezing})
\begin{equation}
    S_2(\zeta)\ket{p, q} = \sum_{k=0}^{\infty} \sum_{n=0}^{\min{(p,q)}}\!\!\mathcal{A}^{(2)}_{k,n}(p,q;\zeta)\ket{p\!-\!n\!+\!k, q\!-\!n\!+\!k}, \label{eq:s2fock}
\end{equation}
where $\zeta=re^{i\theta}$, and for the single-mode case, 
\begin{equation}
S_1(\zeta) \ket p = \sum_{k=0}^\infty \sum_{n=0}^{\lfloor p/2\rfloor}\!\mathcal{A}^{(1)}_{k,n}(p;\zeta)\ket{p\!+\!2(k\!-\!n)}. \label{eq:s1fock}
\end{equation}

\begin{figure}[b]
    \centering
    \includegraphics[width=.95\linewidth] {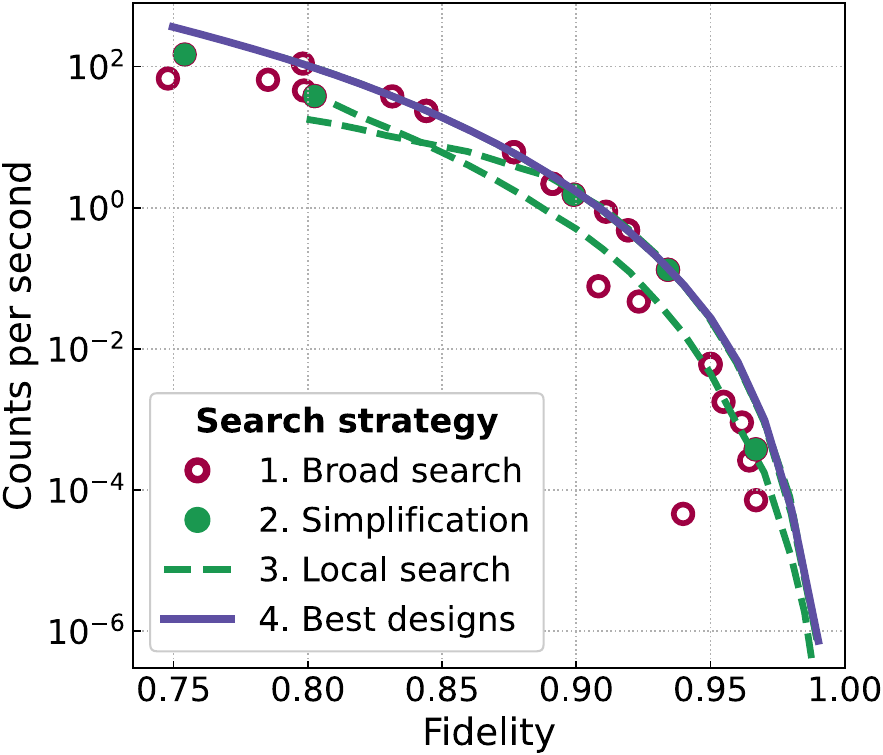}
    \caption{\textbf{Optimization strategy.}
    The search takes four steps:
    \textbf{1)} experiment optimization for random sortings and target fidelities $f_0$ (only a few shown),
    \textbf{2)} simplification of the best candidates computed with a higher photon cutoff (convergence check),
    \textbf{3)} optimization of previous solutions slowly varying $f_0$,
    and \textbf{4)} final choice of best-performing designs.}
    \label{fig:search}
\end{figure}

\begin{figure*}[t]
    \centering
    \includegraphics[width=\linewidth]{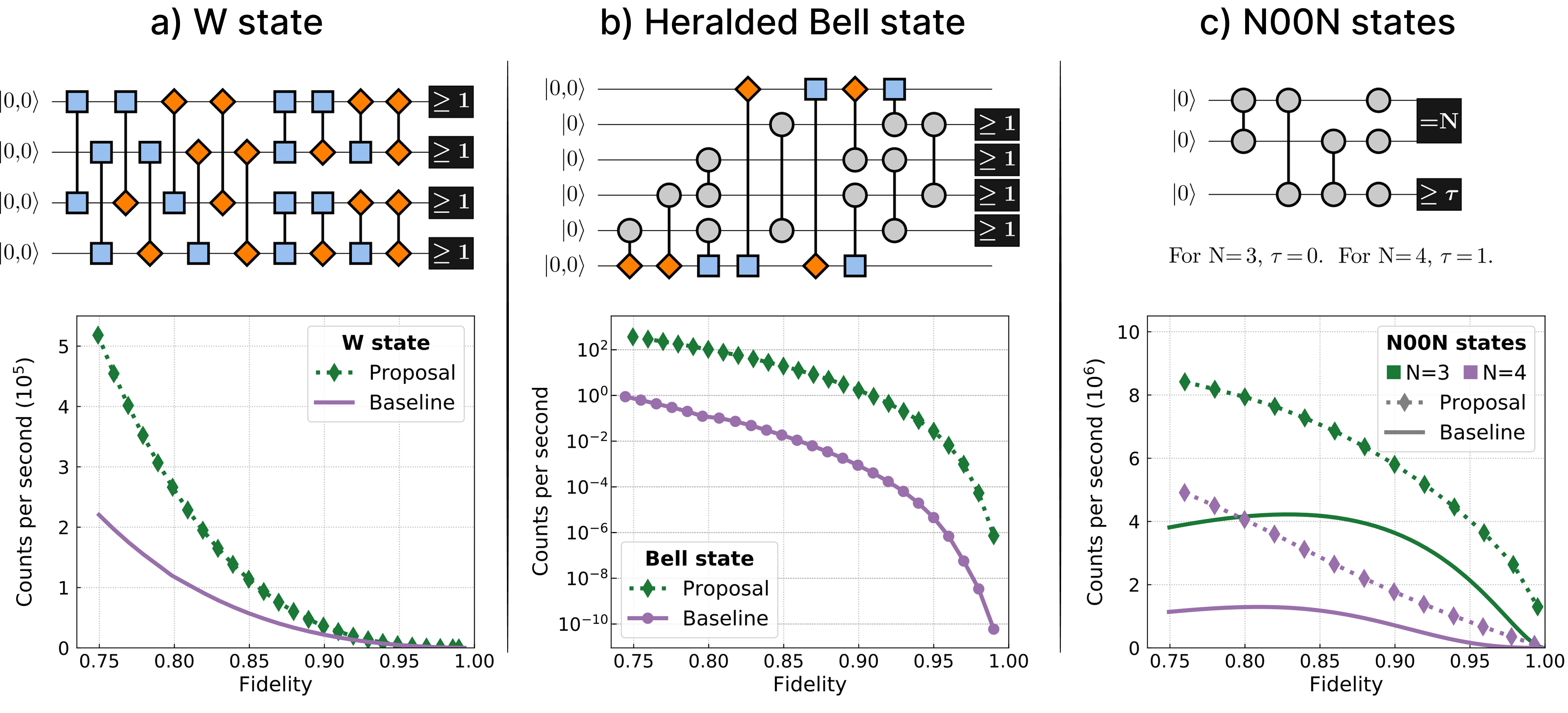}
    \caption{\textbf{High probability entanglement by path identity.} The presented experiments use SPDC crystals (squeeze operators) to generate entanglement. These path identity experiments are compared with previous designs, originally proposed for the low-gain regime, that neglected higher-order terms~\cite{graphs3,sliwa2003conditional,ruiz2023digital}. \textbf{a)} The first design produces a postselected W state of 4 qubits using threshold detectors. \textbf{b)} The second produces a heralded Bell state using 4 ancillary paths, each with a threshold detector. The entangled photons are not detected. \textbf{c)} The last one can generate two variants of N00N states using different parameters and postselection rules. For N=3, the threshold detector is not needed, i.e. $\tau=0$, and the extra photons are lost. For N=4, we use a threshold detector and a single-mode squeezer acting on the ancilla mode.}
\label{fig:summary}
\end{figure*}

\mysection{Automated experimental design}
We want to produce entangled states with high success probability via automated design of quantum experiments \cite{krenn2016automated,krenn2020computer}. 
We know from the baseline experiments that high fidelity can be reached, but not which probabilities to expect.
Thus, we aim for a target fidelity, $f_0$, minimize the loss
\begin{align}
    \mathcal L_\pi({\mathbf r},\bm{\theta})\! =\!& -w_1\log \mathcal{P} + w_2\log\left( (\mathcal{F} - f_0)^2\right)\nonumber\\
    &+ w_3\:||\mathbf{r}||_1+w_4||\bm{\theta}||_1
    +w_5\:\mathcal E,
    \label{eq:loss_function}
\end{align}
where $\mathcal{P}$ is the success probability, $\mathcal{F}$ the fidelity, $\mathbf r\equiv\{r_j\}$ and $\bm \theta \equiv\{\theta_j\}$ the squeezing parameters ($r_je^{\theta_j}=\zeta_j$), and $\pi$ a fixed ordering of the sources. 
The remaining terms are optional: L1 regularization, $||\cdot||_1$, to enforce sparsity, and the truncation error, $\mathcal E = 1 - \text{Tr}(\rho)$, to penalize the loss of amplitudes due to the cutoff on the Fock space.
Sweeping the target fidelity $f_0$ from 75\% to 99\%, we find the Pareto front of the design.
Since $\mathcal P$ may fluctuate by orders of magnitude, the weights, $w_i$, must be tuned for different detection schemes and target fidelities $f_0$. 

Each experiment starts from a multi-mode vacuum, on which we apply a series of squeeze operators in a given order $\pi$: $\{S_j(\zeta_j)\}_\pi$.
The optical paths are then measured by threshold detectors, photon-number resolving (PNR) detectors, or left undetected.
A successful detection is described as a projection $M$ on the resulting state $\rho(\bm \zeta,\pi)$, which gives the success probability: $\mathcal P=\text{Tr}(M\rho)$.
To translate the success probability into expected counts per second, we assume the sources are pumped by a laser source with a repetition rate of 10ns. 
Finally, we trace out the ancillary modes from the state $\rho'\!=\!M\rho M^\dagger / \mathcal P$ and compute the fidelity: $\mathcal F=\bra{\psi}\text{Tr}_\text{anc}(\rho')\ket{\psi}$.

To find the best designs, we randomly order the sources and optimize the parameters $\bm\zeta$.% via automatic differentiation.
We repeat many times for different orderings and target fidelities.
The best candidates for each $f_0$ are recomputed at a higher photon cutoff, to ensure $\mathcal E \to0$, and pruned of redundant sources.
This produces a pool of distinct topologies.
Then, we reoptimize each of them starting from the previously optimized squeezing parameters, now with less sources, for the same target fidelity $f_0$.
Slowly varying $f_0$ across the desired range, we restart the optimizations from the previous optimal parameters.
Selecting the best-performing topology, we obtain a set of experiments characterized by a high-dimensional curve $\bm\zeta(f_0)$.
This offers a simpler implementation than the initial zoo of unalike designs.
Fig.~\ref{fig:search} illustrates the procedure.

Compared to earlier work on automated experiment design~\cite{krenn2021conceptual, ruiz2023digital}, here the source ordering is a search variable.
In the low-gain regime, postselection leaves each path populated by at most one creation operator (see Eq.~\eqref{eq:lowgainGHZ}), so sources effectively commute and operators acting on the same modes merge: $(1 + \zeta_1 a^\dagger b^\dagger)(1 + \zeta_2 a^\dagger b^\dagger) \approx 1 + (\zeta_1 + \zeta_2) a^\dagger b^\dagger$.
These simplifications enable graph-based representations of quantum experiments~\cite{graphs1, graphs2, graphs3}.
Otherwise, squeeze operators commute only when acting on different modes (as in Fig.~\ref{fig:ghz4circuits}b) or, on shared modes, when $\zeta_1 \zeta_2^* = \zeta_1^* \zeta_2$.
Thus, the final state generally depends on the sources' ordering (see N00N baselines in App.~\ref{app:bl_noon}), turning experimental design into a joint search over a discrete ordering and a continuous parameter space.

\mysection{Results}
To showcase our approach, we present new setups to generate diverse forms of entanglement using different detection schemes:
a postselected 4-qubit W state, a heralded Bell state with 4 ancillas, and 2-mode N00N states with 3 and 4 photons, both with an ancillary mode.
We benchmark our designs against previous proposals~\cite{graphs3,sliwa2003conditional,ruiz2023digital}, recomputed with higher pumps.
For a fair comparison, we also optimized the source sorting when required, and adjust the linear optical elements for the heralded Bell (see App.~\ref{app:baselines}). Fig.~\ref{fig:summary} shows the new designs and the performance improvement.

We start with a postselected 4-photon W state $\ket \psi = (\ket{\text{HHHV}}+\ket{\text{HHVH}}+ \ket{\text{HVHH}}+\ket{\text{VHHH}})/2 $, an entangled state robust against photon loss and widely used in quantum information tasks~\cite{dur2000three, agrawal2006perfect}.
The baseline is an early work on path identity with a common squeezing factor for all sources~\cite{graphs3}, without interfering terms.
Our design uses a different topology with additional sources, some of them producing pairs of vertical photons, and distinct phases that enable destructive interference.
These features are deemed redundant in the low-gain regime, and yet they improve the fidelity-probability tradeoff at higher pumps.

Our second experiment generates a Bell state, $\ket \psi = (\ket{\text{HH}}+\ket{\text{VV}})/\sqrt 2$, the most famous form of entanglement~\cite{bell1964einstein, horodecki2009entanglement}. 
The state's generation is conditioned on the detection of 4 ancillary modes that herald successful generation in the unmeasured paths~\cite{forbes2025heralded}.
The baseline combines a type-II SPDC source with several (polarizing) beam splitters and a half-wave plate~\cite{sliwa2003conditional,wagenknecht2010experimental}. 
We fit the system into an 8-mode setup with 4 threshold detectors, like ours, replaced the linear optical elements with equivalent beam splitters, and optimized the parameters to obtain the curve shown in Fig.~\ref{fig:summary}b.
After the ``translation'' and despite the optimization efforts (details in App.~\ref{app:bl_bell}), our setup outperformed the previous design by a wide margin.

The two remaining experiments produce 2-mode N00N states $\ket{\psi} = (\ket{\text{N};0} +\ket{0;\text{N}})/\sqrt{2}$,
with 3 and 4 photons, both using an ancillary mode.
Such states are widely used in metrology~\cite{holland1993interferometric, deng2024quantum}.
In contrast to the polarization-encoded states above, the entanglement here is between photon numbers in the two main modes, which a PNR detector pins to N.
The constraints on the ancillary path are more lenient.
For N=3, photon-pair creation guarantees an odd ancillary photon count, so no detector is needed.
For N=4, a single threshold detector verifies that extra photons are in the ancillary path, at least two if the detector clicks.
Two photons reaching a single threshold detector was not considered by the baseline. 
Yet, as it improves the performance, we applied the same trick for comparison (see App.~\ref{app:bl_noon}), and our designs still outperform it.
For N=3, we use the same sources as the baseline but different squeezing parameters.
This led to new interference patterns providing a better performance.

\begin{figure}[t]
    \centering
    \includegraphics[width=.95\linewidth]{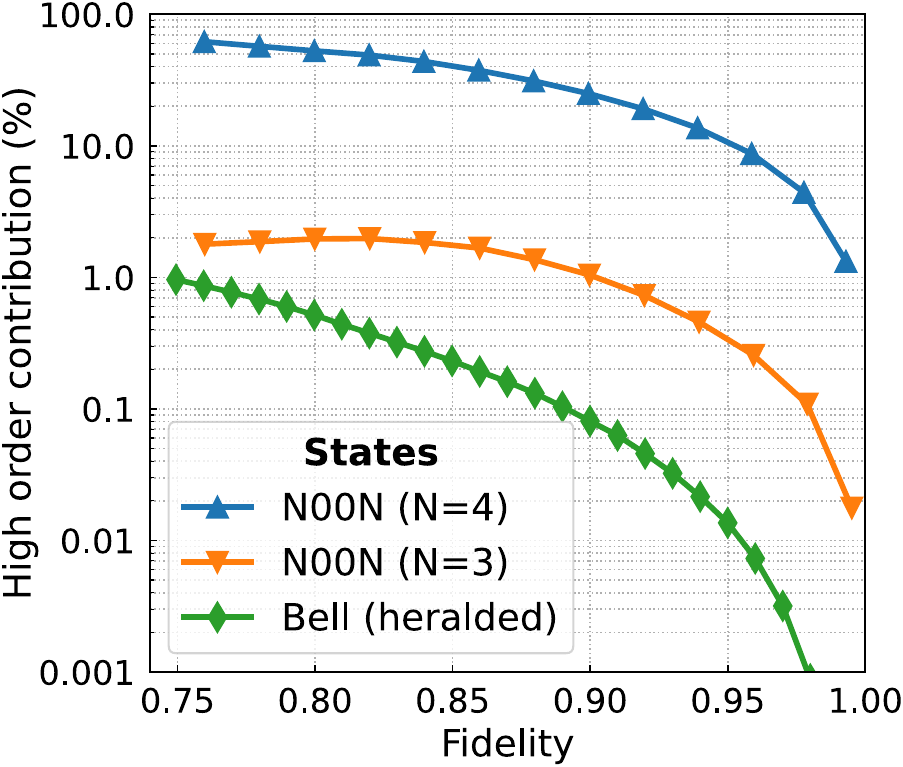}
    \caption{\textbf{Relative contribution to the fidelity of high-order terms.} 
    When using ancilla paths, creation events with a higher photon-number can also lead to the target (reduced) state.
    Yet, lower-order terms are still predominant.} \label{fig:hot}
\end{figure}

Beyond the improvements shown in Fig.~\ref{fig:summary}, the last 3 experiments share a common feature: their fidelity benefits from contributions with different photon numbers in the ancillary paths.
As shown in Fig.~\ref{fig:hot}, these effects are larger for the N00N states.
For N=4, at 90\% fidelity, more than a fifth comes from high photon-number emissions.
Perturbative approaches overlook such terms, our framework turns them into a resource.

\mysection{Discussion}
By computing emission events with an arbitrary number of photons, we designed efficient experiments for entanglement generation.
Our algorithm combined the Fock-basis description of SPDC sources with different detection schemes.
Aided by automatic differentiation, we optimized path identity setups to maximize fidelity and success probability.
We surpassed previous proposals to produce a postselected W state, a heralded Bell state, and two N00N states.
Moreover, when using ancillary paths, the fidelity benefited from multi-photon events, an unexplored resource in the low-gain regime. 

The generated states were diverse and broadly applied.
Yet, as the Hilbert space grew exponentially with the number of modes, we modeled relatively small systems.
While the Gaussian formalism enables a more compact representation of our experiments~\cite{braunstein2005quantum, weedbrook2012gaussian}, symplectic transformations cannot model arbitrary Fock states, and the extraction of Fock-state amplitudes via Hafnians is computationally demanding~\cite{hamilton2017gaussian, killoran2019strawberry, gupt2019walrus, mrmustard2024}.
Scaling the approach to larger systems for arbitrary initial states may require new simulation techniques.

Our algorithm currently optimizes squeezing parameters after randomizing the orderings of the non-commuting SPDC sources.
While symmetries reduce the count, we still face a large discrete search space and must resort to expensive optimizations without a guarantee of global optimality. Theoretical intuition or rigorous design rules for the experiments' topology are much needed.
Related questions were explored in the graph-based representation of quantum experiments~\cite{krenn2019questions}, showing how graph theory can guide experimental design~\cite{vardi2023solving, chandran2024krenn, Chandran2024graphtheoretic}.
It would be interesting to develop analogous principles for the non-commuting, high-gain regime considered here.

Finally, while our proposals rely primarily on SPDC sources, linear optical elements can be included, as shown in the Bell state baseline.
Including such elements as part of the optimization, alongside new non-linear tools, would expand the accessible design space.
Furthermore, different hardware platforms also impose their own constraints.
To name a few: photon losses, detector inefficiencies, or structural limits in integrated photonic circuits.
Encoding these into the design process can foster the adoption of new experimental solutions.

\mysection{Acknowledgments}
The authors thank Soeren Arlt, Kai Wang, and Xiaosong Ma for useful discussions.
MK acknowledges support by the European Research Council (ERC) under the European Union’s Horizon Europe research and innovation programme (ERC-2024-STG, 101165179, ArtDisQ) and from the German Research Foundation DFG (EXC 2064/1, Project 390727645).

\bibliography{refs}

@article{horodecki2009entanglement,
  title = {Quantum entanglement},
  author = {Horodecki, Ryszard and Horodecki, Pawe\l{} and Horodecki, Micha\l{} and Horodecki, Karol},
  journal = {Rev. Mod. Phys.},
  volume = {81},
  issue = {2},
  pages = {865--942},
  numpages = {0},
  year = {2009},
  month = {Jun},
  publisher = {American Physical Society},
  doi = {10.1103/RevModPhys.81.865},
  url = {https://link.aps.org/doi/10.1103/RevModPhys.81.865}
}

@article{bourassa2021blueprint,
  title={Blueprint for a scalable photonic fault-tolerant quantum computer},
  author={Bourassa, J Eli and Alexander, Rafael N and Vasmer, Michael and Patil, Ashlesha and Tzitrin, Ilan and Matsuura, Taknori and Su, Daiqin and Baragiola, Ben Q and Gupt, Brajesh and Sabapathy, Krishna K and others},
  journal={Quantum},
  volume={5},
  pages={392},
  year={2021},
  publisher={Verein zur F{\"o}rderung des Open Access Publizierens in den Quantenwissenschaften},
  url={https://doi.org/10.22331/q-2021-02-04-392}
}

@article{xanadu2025scaling,
  title={Scaling and networking a modular photonic quantum computer},
  author={Aghaee Rad, H and Ainsworth, Thomas and Alexander, Rafael N and Altieri, Brandon and Askarani, Mohsen F and Baby, R and Banchi, Leonardo and Baragiola, Ben Q and Bourassa, J Eli and Chadwick, RS and others},
  journal={Nature},
  volume={638},
  number={8052},
  pages={912--919},
  year={2025},
  publisher={Nature Publishing Group UK London},
  url={https://doi.org/10.1038/s41586-024-08406-9}}

@article{psiquantum2025,
  title={A manufacturable platform for photonic quantum computing},
  author={Psi{Q}uantum},
  journal={Nature},
  volume={641},
  number={8064},
  pages={876--883},
  year={2025},
  publisher={Nature Publishing Group UK London},
  url={https://doi.org/10.1038/s41586-025-08820-7}}

@article{huang2024entanglement,
  title={Entanglement-enhanced quantum metrology: From standard quantum limit to {H}eisenberg limit},
  author={Huang, Jiahao and Zhuang, Min and Lee, Chaohong},
  journal={Applied Physics Reviews},
  volume={11},
  number={3},
  year={2024},
  publisher={AIP Publishing},
  url={https://doi.org/10.1063/5.0204102}
}

@article{ma2025unraveling,
  title={Unraveling quantum phase estimation: exploring the impact of multi-photon interference on the quantum Fisher information},
  author={Ma, Annameng and Magnoni, Agustina G and Larotonda, Miguel A and Knoll, Laura T},
  journal={Quantum Science and Technology},
  volume={10},
  number={3},
  pages={035021},
  year={2025},
  publisher={IOP Publishing},
  url={https://doi.org/10.1088/2058-9565/add04d}
}

@article{azuma2023repeaters,
  title = {Quantum repeaters: From quantum networks to the quantum internet},
  author = {Azuma, Koji and Economou, Sophia E. and Elkouss, David and Hilaire, Paul and Jiang, Liang and Lo, Hoi-Kwong and Tzitrin, Ilan},
  journal = {Rev. Mod. Phys.},
  volume = {95},
  issue = {4},
  pages = {045006},
  numpages = {66},
  year = {2023},
  month = {Dec},
  publisher = {American Physical Society},
  url={https://doi.org/10.1103/RevModPhys.95.045006}
}

@article{zhang2025towards,
  title={Towards global quantum key distribution},
  author={Zhang, Haoran and Zhu, Haotao and He, Ruihua and Zhang, Yan and Ding, Chao and Hanzo, Lajos and Gao, Weibo},
  journal={Nature Reviews Electrical Engineering},
  pages={1--13},
  year={2025},
  publisher={Nature Publishing Group UK London},
  url={https://doi.org/10.1038/s44287-025-00238-7},  
}

@article{luo2023recent,
  title={Recent progress in quantum photonic chips for quantum communication and internet},
  author={Luo, Wei and Cao, Lin and Shi, Yuzhi and Wan, Lingxiao and Zhang, Hui and Li, Shuyi and Chen, Guanyu and Li, Yuan and Li, Sijin and Wang, Yunxiang and others},
  journal={Light: Science \& Applications},
  volume={12},
  number={1},
  pages={175},
  year={2023},
  publisher={Nature Publishing Group UK London},
url={https://doi.org/10.1038/s41377-023-01173-8}}

@article{wang2025scalable,
  title={Scalable photonic quantum technologies},
  author={Wang, Hui and Ralph, Timothy C and Renema, Jelmer J and Lu, Chao-Yang and Pan, Jian-Wei},
  journal={Nature Materials},
  volume={24},
  number={12},
  pages={1883--1897},
  year={2025},
  publisher={Nature Publishing Group UK London},
url={https://doi.org/10.1038/s41563-025-02306-7}}

@article{vanleent2022entangling,
  title={Entangling single atoms over 33 km telecom fibre},
  author={van Leent, Tim and Bock, Matthias and Fertig, Florian and Garthoff, Robert and Eppelt, Sebastian and Zhou, Yiru and Malik, Pooja and Seubert, Matthias and Bauer, Tobias and Rosenfeld, Wenjamin and others},
  journal={Nature},
  volume={607},
  number={7917},
  pages={69--73},
  year={2022},
  publisher={Nature Publishing Group},
  url={https://doi.org/10.1038/s41586-022-04764-4}
}

@article{krvzivc2023towards,
  title={Towards metropolitan free-space quantum networks},
  author={Kr{\v{z}}i{\v{c}}, Andrej and Sharma, Sakshi and Spiess, Christopher and Chandrashekara, Uday and T{\"o}pfer, Sebastian and Sauer, Gregor and Gonz{\'a}lez-Mart{\'\i}n del Campo, Luis Javier and Kopf, Teresa and Petscharnig, Stefan and Grafenauer, Thomas and others},
  journal={npj Quantum Information},
  volume={9},
  number={1},
  pages={95},
  year={2023},
  publisher={Nature Publishing Group UK London},
  url={https://doi.org/10.1038/s41534-023-00754-0}}

@article{kucera2024demonstration,
  title={Demonstration of quantum network protocols over a 14-km urban fiber link},
  author={Kucera, Stephan and Haen, Christian and Arensk{\"o}tter, Elena and Bauer, Tobias and Meiers, Jonas and Sch{\"a}fer, Marlon and Boland, Ross and Yahyapour, Milad and Lessing, Maurice and Holzwarth, Ronald and others},
  journal={npj Quantum Information},
  volume={10},
  number={1},
  pages={88},
  year={2024},
  publisher={Nature Publishing Group UK London},
  url={https://doi.org/10.1038/s41534-024-00886-x}}

@article{pittaluga2025long,
  title={Long-distance coherent quantum communications in deployed telecom networks},
  author={Pittaluga, Mirko and Lo, Yuen San and Brzosko, Adam and Woodward, Robert I and Scalcon, Davide and Winnel, Matthew S and Roger, Thomas and Dynes, James F and Owen, Kim A and Ju{\'a}rez, Sergio and others},
  journal={Nature},
  volume={640},
  number={8060},
  pages={911--917},
  year={2025},
  publisher={Nature Publishing Group UK London},
url={https://doi.org/10.1038/s41586-025-08801-w}}

@article{nielsen2025programmable,
  title={Programmable nonlinear quantum photonic circuits},
  author={Nielsen, Kasper H and Wang, Ying and Deacon, Edward CR and Sund, Patrik I and Liu, Zhe and Scholz, Sven and Wieck, Andreas D and Ludwig, Arne and Midolo, Leonardo and S{\o}rensen, Anders S and others},
  journal={Nature Communications},
  year={2025},
  publisher={Nature Publishing Group UK London},
  url={https://doi.org/10.1038/s41467-025-66205-w}
}

@article{ruiz2023digital,
  title={Digital discovery of 100 diverse quantum experiments with {P}yTheus},
  author={Ruiz-Gonzalez, Carlos and Arlt, S{\"o}ren and Petermann, Jan and Sayyad, Sharareh and Jaouni, Tareq and Karimi, Ebrahim and Tischler, Nora and Gu, Xuemei and Krenn, Mario},
  journal={Quantum},
  volume={7},
  pages={1204},
  year={2023},
  publisher={Verein zur F{\"o}rderung des Open Access Publizierens in den Quantenwissenschaften},
  url={https://doi.org/10.22331/q-2023-12-12-1204}
}

@article{forbes2025heralded,
  title={Heralded generation of entanglement with photons},
  author={Forbes, Imogen and Ghafari, Farzad and Deacon, Edward CR and Singh, Sukhjit P and Lavie, Emilien and Yard, Patrick and Shaw, Reece D and Laing, Anthony and Tischler, Nora},
  journal={Reports on Progress in Physics},
  volume={88},
  number={8},
  pages={086002},
  year={2025},
  publisher={IOP Publishing},
  url={https://doi.org/10.1088/1361-6633/adf85e}}

@article{chizhov1993photon,
  title={Photon statistics and phase properties of two-mode squeezed number states},
  author={Chizhov, AV and Murzakhmetov, BK},
  journal={Physics Letters A},
  volume={176},
  number={1-2},
  pages={33--40},
  year={1993},
  publisher={Elsevier},
  url={https://doi.org/10.1016/0375-9601(93)90312-N}
}

@article{quesada2022beyond,
  title={Beyond photon pairs—nonlinear quantum photonics in the high-gain regime: a tutorial},
  author={Quesada, Nicol{\'a}s and Helt, LG and Menotti, Matteo and Liscidini, Marco and Sipe, JE},
  journal={Advances in Optics and Photonics},
  volume={14},
  number={3},
  pages={291--403},
  year={2022},
  publisher={Optica Publishing Group},
  url={https://doi.org/10.1364/AOP.445496}
}

@article{bulmer2022threshold,
  title = {Threshold detection statistics of bosonic states},
  author = {Bulmer, J. F. F. and Paesani, S. and Chadwick, R. S. and Quesada, N.},
  journal = {Phys. Rev. A},
  volume = {106},
  issue = {4},
  pages = {043712},
  numpages = {14},
  year = {2022},
  month = {Oct},
  publisher = {American Physical Society},
  url={https://doi.org/10.1103/PhysRevA.106.043712}
}

@article{chinni2024beyond,
  title = {Beyond the parametric approximation: Pump depletion, entanglement, and squeezing in macroscopic down-conversion},
  author = {Chinni, Karthik and Quesada, Nicol\'as},
  journal = {Phys. Rev. A},
  volume = {110},
  issue = {1},
  pages = {013712},
  numpages = {21},
  year = {2024},
  month = {Jul},
  publisher = {American Physical Society},
  url={https://doi.org/10.1103/PhysRevA.110.013712}
}

@article{gu2025analytical,
  title={Analytical {F}ock Representation of Two-Mode Squeezing for Quantum Interference},
  author={Gu, Xuemei and Ruiz-Gonzalez, Carlos and Krenn, Mario},
  journal={arXiv:2511.16529},
  year={2025},
  url={https://arxiv.org/abs/2511.16529}
}

@article{graphs1,
  title = {Quantum Experiments and Graphs: Quantum States as Coherent Superpositions of Perfect Matchings},
  author = {Krenn, Mario and Gu, Xuemei and Zeilinger, Anton},
  journal = {Phys. Rev. Lett.},
  volume = {119},
  issue = {24},
  pages = {240403},
  numpages = {6},
  year = {2017},
  month = {Dec},
  publisher = {American Physical Society},
url={https://doi.org/10.1103/PhysRevLett.119.240403}
}

@article{graphs2,
  title={Quantum experiments and graphs {II}: Quantum interference, computation, and state generation},
  author={Gu, Xuemei and Erhard, Manuel and Zeilinger, Anton and Krenn, Mario},
  journal={Proceedings of the National Academy of Sciences},
  volume={116},
  number={10},
  pages={4147--4155},
  year={2019},
  publisher={National Academy of Sciences},
url={https://doi.org/10.1073/pnas.1815884116}}

@article{graphs3,
  title = {Quantum experiments and graphs {III}: High-dimensional and multiparticle entanglement},
  author = {Gu, Xuemei and Chen, Lijun and Zeilinger, Anton and Krenn, Mario},
  journal = {Phys. Rev. A},
  volume = {99},
  issue = {3},
  pages = {032338},
  numpages = {11},
  year = {2019},
  month = {Mar},
  publisher = {American Physical Society},
  url = {https://doi.org/10.1103/PhysRevA.99.032338}
}

@article{krenn2017pathidentity,
  title = {Entanglement by Path Identity},
  author = {Krenn, Mario and Hochrainer, Armin and Lahiri, Mayukh and Zeilinger, Anton},
  journal = {Phys. Rev. Lett.},
  volume = {118},
  issue = {8},
  pages = {080401},
  numpages = {6},
  year = {2017},
  month = {Feb},
  publisher = {American Physical Society},
  url = {https://link.aps.org/doi/10.1103/PhysRevLett.118.080401}
}

@article{zou91induced,
  title = {Induced coherence and indistinguishability in optical interference},
  author = {Zou, X. Y. and Wang, L. J. and Mandel, L.},
  journal = {Phys. Rev. Lett.},
  volume = {67},
  issue = {3},
  pages = {318--321},
  numpages = {0},
  year = {1991},
  month = {Jul},
  publisher = {American Physical Society},
url={https://doi.org/10.1103/PhysRevLett.67.318}}

@article{wang91induced,
  title = {Induced coherence without induced emission},
  author = {Wang, L. J. and Zou, X. Y. and Mandel, L.},
  journal = {Phys. Rev. A},
  volume = {44},
  issue = {7},
  pages = {4614--4622},
  numpages = {0},
  year = {1991},
  month = {Oct},
  publisher = {American Physical Society},
url={https://doi.org/10.1103/PhysRevA.44.4614}}

@article{kysela2020path,
author = {Jaroslav Kysela  and Manuel Erhard  and Armin Hochrainer  and Mario Krenn  and Anton Zeilinger },
title = {Path identity as a source of high-dimensional entanglement},
journal = {Proceedings of the National Academy of Sciences},
volume = {117},
number = {42},
pages = {26118-26122},
year = {2020},
url={https://doi.org/10.1073/pnas.2011405117}}

@article{hochrainer2022quantum,
  title={Quantum indistinguishability by path identity and with undetected photons},
  author={Hochrainer, Armin and Lahiri, Mayukh and Erhard, Manuel and Krenn, Mario and Zeilinger, Anton},
  journal={Reviews of Modern Physics},
  volume={94},
  number={2},
  pages={025007},
  year={2022},
  publisher={APS},
url={https://doi.org/10.1103/RevModPhys.94.025007}
}

@article{barreto2022quantum,
  title={Quantum imaging and metrology with undetected photons: tutorial},
  author={Barreto Lemos, Gabriela and Lahiri, Mayukh and Ramelow, Sven and Lapkiewicz, Radek and Plick, William N},
  journal={Journal of the Optical Society of America B},
  volume={39},
  number={8},
  pages={2200--2228},
  year={2022},
  publisher={Optica Publishing Group},
url={https://doi.org/10.1364/JOSAB.456778}
}

@article{wang2024entangling,
  title = {Entangling Independent Particles by Path Identity},
  author = {Wang, Kai and Hou, Zhaohua and Qian, Kaiyi and Chen, Leizhen and Krenn, Mario and Zhu, Shining and Ma, Xiao-Song},
  journal = {Phys. Rev. Lett.},
  volume = {133},
  issue = {23},
  pages = {233601},
  numpages = {6},
  year = {2024},
  month = {Dec},
  publisher = {American Physical Society},
url={https://doi.org/10.1103/PhysRevLett.133.233601}
}

@article{hu2025observation,
  title={Observation of Genuine High-dimensional Multi-partite Non-locality in Entangled Photon States},
  author={Hu, Xiao-Min and Huang, Cen-Xiao and d'Alessandro, Nicola and Cobucci, Gabriele and Zhang, Chao and Guo, Yu and Huang, Yun-Feng and Li, Chuan-Feng and Guo, Guang-Can and Gao, Xiaoqin and Huber, Marcus and Tavakoli, Armin and Liu, Bi-Heng},
  journal={Nature Communications},
  volume={16},
  number={1},
  pages={5017},
  year={2025},
  publisher={Nature Publishing Group UK London},
  url={https://doi.org/10.1038/s41467-025-59717-y}
}

@article{bernecker2025quantum,
  title={Engineering of maximally entangled orbital angular momentum states via path identity},
  author={Bernecker, Richard and Baghdasaryan, Baghdasar and Fritzsche, Stephan},
  journal={Physical Review A},
  volume={112},
  number={6},
  pages={063701},
  year={2025},
  publisher={APS},
  url = {https://link.aps.org/doi/10.1103/9qrm-chgg}
}

@article{dasgupta1996disentanglement,
  title={Disentanglement formulas: An alternative derivation and some applications to squeezed coherent states},
  author={DasGupta, Ananda},
  journal={American Journal of Physics},
  volume={64},
  number={11},
  pages={1422--1427},
  url={https://doi.org/10.1119/1.18183},
  year={1996}
}

@article{krenn2016automated,
  title={Automated search for new quantum experiments},
  author={Krenn, Mario and Malik, Mehul and Fickler, Robert and Lapkiewicz, Radek and Zeilinger, Anton},
  journal={Physical Review Letters},
  volume={116},
  number={9},
  pages={090405},
  year={2016},
  publisher={APS},
url={https://doi.org/10.1103/PhysRevLett.116.090405}
}

@article{krenn2020computer,
  title={Computer-inspired quantum experiments},
  author={Krenn, Mario and Erhard, Manuel and Zeilinger, Anton},
  journal={Nature Reviews Physics},
  volume={2},
  number={11},
  pages={649--661},
  year={2020},
  publisher={Nature Publishing Group UK London},
  url = {https://doi.org/10.1038/s42254-020-0230-4}
}

@article{krenn2021conceptual,
  title = {Conceptual Understanding through Efficient Automated Design of Quantum Optical Experiments},
  author = {Krenn, Mario and Kottmann, Jakob S. and Tischler, Nora and Aspuru-Guzik, Al\'an},
  journal = {Phys. Rev. X},
  volume = {11},
  issue = {3},
  pages = {031044},
  numpages = {15},
  year = {2021},
  month = {Aug},
  publisher = {American Physical Society},
  doi = {10.1103/PhysRevX.11.031044},
  url = {https://doi.org/10.1103/PhysRevX.11.031044}
}

@article{dur2000three,
  title = {Three qubits can be entangled in two inequivalent ways},
  author = {D\"ur, W. and Vidal, G. and Cirac, J. I.},
  journal = {Phys. Rev. A},
  volume = {62},
  issue = {6},
  pages = {062314},
  numpages = {12},
  year = {2000},
  month = {Nov},
  publisher = {American Physical Society},
  url={https://doi.org/10.1103/PhysRevA.62.062314}
}

@article{agrawal2006perfect,
  title = {Perfect teleportation and superdense coding with {W} states},
  author = {Agrawal, Pankaj and Pati, Arun},
  journal = {Phys. Rev. A},
  volume = {74},
  issue = {6},
  pages = {062320},
  numpages = {5},
  year = {2006},
  month = {Dec},
  publisher = {American Physical Society},
  url={https://doi.org/10.1103/PhysRevA.74.062320}
}

@article{bell1964einstein,
  title={On the {E}instein {P}odolsky {R}osen paradox},
  author={Bell, John S},
  journal={Physics Physique Fizika},
  volume={1},
  number={3},
  pages={195},
  year={1964},
  publisher={APS},
  url={https://doi.org/10.1103/PhysicsPhysiqueFizika.1.195}
}

@article{sliwa2003conditional,
  title={Conditional preparation of maximal polarization entanglement},
  author={{\'S}liwa, Cezary and Banaszek, Konrad},
  journal={Physical Review A},
  volume={67},
  number={3},
  pages={030101},
  year={2003},
  publisher={APS},
  url={https://doi.org/10.1103/PhysRevA.67.030101}
}

@article{wagenknecht2010experimental,
  title={Experimental demonstration of a heralded entanglement source},
  author={Wagenknecht, Claudia and Li, Che-Ming and Reingruber, Andreas and Bao, Xiao-Hui and Goebel, Alexander and Chen, Yu-Ao and Zhang, Qiang and Chen, Kai and Pan, Jian-Wei},
  journal={Nature Photonics},
  volume={4},
  number={8},
  pages={549--552},
  year={2010},
  publisher={Nature Publishing Group UK London},
  url={https://doi.org/10.1038/nphoton.2010.123}
}

@article{braunstein2005quantum,
  title={Quantum information with continuous variables},
  author={Braunstein, Samuel L and Van Loock, Peter},
  journal={Reviews of modern physics},
  volume={77},
  number={2},
  pages={513--577},
  year={2005},
  publisher={APS},
url={https://doi.org/10.1103/RevModPhys.77.513}
}

@article{weedbrook2012gaussian,
  title = {Gaussian quantum information},
  author = {Weedbrook, Christian and Pirandola, Stefano and Garc\'{\i}a-Patr\'on, Ra\'ul and Cerf, Nicolas J. and Ralph, Timothy C. and Shapiro, Jeffrey H. and Lloyd, Seth},
  journal = {Rev. Mod. Phys.},
  volume = {84},
  issue = {2},
  pages = {621--669},
  numpages = {0},
  year = {2012},
  publisher = {American Physical Society},
  url = {https://doi.org/10.1103/RevModPhys.84.621}
}

@article{hamilton2017gaussian,
  title = {Gaussian Boson Sampling},
  author = {Hamilton, Craig S. and Kruse, Regina and Sansoni, Linda and Barkhofen, Sonja and Silberhorn, Christine and Jex, Igor},
  journal = {Phys. Rev. Lett.},
  volume = {119},
  issue = {17},
  pages = {170501},
  numpages = {5},
  year = {2017},
  publisher = {American Physical Society},
  url = {https://doi.org/10.1103/PhysRevLett.119.170501}
}

@article{killoran2019strawberry,
  title={Strawberry fields: A software platform for photonic quantum computing},
  author={Killoran, Nathan and Izaac, Josh and Quesada, Nicol{\'a}s and Bergholm, Ville and Amy, Matthew and Weedbrook, Christian},
  journal={Quantum},
  volume={3},
  pages={129},
  year={2019},
  publisher={Verein zur F{\"o}rderung des Open Access Publizierens in den Quantenwissenschaften},
  url={http://dx.doi.org/10.22331/q-2019-03-11-129},
}

@article{gupt2019walrus,
  title = {The {Walrus}: a library for the calculation of hafnians, {Hermite} polynomials and {Gaussian} boson sampling},
  author = {Gupt, Brajesh and Izaac, Josh and Quesada, Nicol{\'a}s},
  journal = {Journal of Open Source Software},
  volume = {4},
  number = {44},
  pages = {1705},
  year = {2019},
  publisher = {The Open Journal},
  url = {https://doi.org/10.21105/joss.01705}
}

@misc{mrmustard2024,
  author = {{Xanadu Quantum Technologies}},
  title = {{Mr Mustard}: A differentiable bridge between phase space and {F}ock space},
  version = {0.7.3},
  year = {2024},
  url = {https://mrmustard.readthedocs.io/en/stable/},
}

@article{holland1993interferometric,
  title = {Interferometric detection of optical phase shifts at the {H}eisenberg limit},
  author = {Holland, M. J. and Burnett, K.},
  journal = {Phys. Rev. Lett.},
  volume = {71},
  issue = {9},
  pages = {1355--1358},
  numpages = {0},
  year = {1993},
  month = {Aug},
  publisher = {American Physical Society},
  url = {https://doi.org/10.1103/PhysRevLett.71.1355}
}

@article{deng2024quantum,
  title={Quantum-enhanced metrology with large {F}ock states},
  author={Deng, Xiaowei and Li, Sai and Chen, Zi-Jie and Ni, Zhongchu and Cai, Yanyan and Mai, Jiasheng and Zhang, Libo and Zheng, Pan and Yu, Haifeng and Zou, Chang-Ling and others},
  journal={Nature Physics},
  volume={20},
  number={12},
  pages={1874--1880},
  year={2024},
  publisher={Nature Publishing Group UK London},
  url={https://doi.org/10.1038/s41567-024-02619-5}
}

@article{krenn2019questions,
  title={Questions on the Structure of Perfect Matchings Inspired by Quantum Physics},
  author={Krenn, Mario and Gu, Xuemei and Soltesz, Daniel},
  journal={Proceedings of the 2nd Croatian Combinatorial Days},
  year={2019},
  url={https://doi.org/10.5592/CO/CCD.2018.05}
}

@article{vardi2023solving,
  title={Solving quantum-inspired perfect matching problems via Tutte-theorem-based hybrid Boolean constraints},
  author={Vardi, Moshe Y and Zhang, Zhiwei},
  journal={Proceedings of the Thirty-Second International Joint Conference on Artificial Intelligence},
  pages={2039--2048},
  year={2023},
url={https://doi.org/10.24963/ijcai.2023/227}}

@inproceedings{chandran2024krenn,
  title={Krenn-{G}u Conjecture for Sparse Graphs},
  author={Chandran, L Sunil and Gajjala, Rishikesh and Illickan, Abraham M},
  booktitle={49th International Symposium on Mathematical Foundations of Computer Science (MFCS 2024)},
  pages={41--1},
  year={2024},
  organization={Schloss Dagstuhl--Leibniz-Zentrum f{\"u}r Informatik},
  url={https://doi.org/10.4230/LIPIcs.MFCS.2024.41}}

@article{Chandran2024graphtheoretic,
  url = {https://doi.org/10.22331/q-2024-07-03-1396},
  title = {Graph-theoretic insights on the constructability of complex entangled states},
  author = {Chandran, L. Sunil and Gajjala, Rishikesh},
  journal = {{Quantum}},
  issn = {2521-327X},
  publisher = {{Verein zur F{\"{o}}rderung des Open Access Publizierens in den Quantenwissenschaften}},
  volume = {8},
  pages = {1396},
  year = {2024}
}

@article{truax1985baker,
  title={Baker-Campbell-Hausdorff relations and unitarity of {SU}(2) and {SU}(1, 1) squeeze operators},
  author={Truax, D Rodney},
  journal={Physical Review D},
  volume={31},
  number={8},
  pages={1988},
  year={1985},
  publisher={APS},
  url={https://doi.org/10.1103/PhysRevD.31.1988}
}

@article{stoler1970equivalence,
  title = {Equivalence Classes of Minimum Uncertainty Packets},
  author = {Stoler, David},
  journal = {Phys. Rev. D},
  volume = {1},
  issue = {12},
  pages = {3217--3219},
  numpages = {0},
  year = {1970},
  publisher = {American Physical Society},
url={https://doi.org/10.1103/PhysRevD.1.3217},
}

@article{kok2000postselected,
  title = {Postselected versus nonpostselected quantum teleportation using parametric down-conversion},
  author = {Kok, Pieter and Braunstein, Samuel L.},
  journal = {Phys. Rev. A},
  volume = {61},
  issue = {4},
  pages = {042304},
  numpages = {10},
  year = {2000},
  month = {Mar},
  publisher = {American Physical Society},
  url={https://doi.org/10.1103/PhysRevA.61.042304}
}

\newpage
\section*{Appendix}

\subsection{Squeeze operators}\label{app:squeezing}
\subsubsection{Two-mode squeezing}
SPDC sources can be modeled by a squeezer operator,
$S_2(\zeta) = \exp \left(\zeta^* a b  - \zeta a^\dagger b^\dagger\right)$, acting over 2-mode Fock states.
$\zeta = r e^{i\theta}$ is the complex squeezing parameter, and $a$ ($a^\dagger$) and $b$ ($b^\dagger$) are the annihilation (creation) operators of the 2 modes.
Using the disentanglement formula for the $\mathrm{SU}(1,1)$ Lie algebra~\cite{truax1985baker, dasgupta1996disentanglement}, the squeezer operator can be rewritten in its normal-ordered form
\begin{align}
S_2(r,\theta) &=\exp \left(-e^{i\theta} \tanh r \, a^\dagger b^\dagger \right)\nonumber\\
&\quad\times \exp \left( -\ln(\cosh r) \, (a^\dagger a + b^\dagger b + 1) \right)\nonumber\\
&\quad\times \exp \left(e^{-i\theta} \tanh r \, ab \right).\label{eq:S2normal_order}
\end{align}
Applied to a Fock state $\ket{p,q}$, this operator generates~\cite{gu2025analytical}
\begin{align}
\ket{\psi}=&S_2(r,\theta)\ket{p, q}\nonumber\\
=&\sum_{k=0}^{\infty} \sum_{n=0}^{\min{(p,q)}}\frac{(-e^{i\theta}\tanh r)^k(e^{-i\theta} \sinh r\cosh r)^n }{(\cosh r)^{p+q+1}} \nonumber\\
&\times \,\sqrt{\binom{p}{n}\binom{q}{n}\binom{p-n+k}{k}\binom{q-n+k}{k}}\nonumber\\
&\times \,\ket{p-n+k, q-n+k}
\label{eq:S2Fockpq}.
\end{align}
Accordingly, no photon is absorbed when the squeezer acts on \textit{any} vacuum, even if only in one mode
\begin{align}
\ket{\psi}=&S_2(r,\theta)\ket{p, 0}\nonumber\\
=&\sum_{k=0}^{\infty} \frac{(-e^{i\theta}\tanh r)^k}{(\cosh r)^{p+1}}\sqrt{\frac{(p+k)!}{p!\,k!}}\,\ket{p+k, k}. \label{eq:S2Fockvacuum}
\end{align}

\subsubsection{Single-mode squeezing}
While the 2-mode squeezing is our main workhorse, we use single mode squeezing for the N00N states: $S_1(\zeta) = \exp \left(\frac{1}{2}(\zeta^* a^2  - \zeta (a^\dagger)^2)\right)$.
Again, $\zeta = re^{i\theta}$ is the complex squeezing parameter, and $a$ ($a^\dagger$) is the annihilation (creation) operator~\cite{stoler1970equivalence}.
From a similar Lie algebra to that for the 2-mode case, we obtain the normal-ordered form%\footnote{Without the factor $1/2$ of $S_1(\zeta)$ one would replace $r$ with $2r$.}
\begin{align}
S_1(r,\theta) =&\exp \left(\frac{-e^{i\theta} \tanh r \, (a^\dagger)^2}{2} \right)\nonumber \\
&\times\exp \left( -\ln(\cosh r) \,\left( a^\dagger a + \frac{1}{2}\right) \right)\nonumber\\
&\times \exp \left(\frac{e^{-i\theta} \tanh r \, a^2}{2} \right),
\end{align}
which applied to a state $\ket p$, leads to
\begin{align}
\ket{\psi}=& S_1(r,\theta) \ket p \nonumber\\
=&\sum_{k=0}^\infty \sum_{n=0}^{\lfloor p/2\rfloor}\frac{(-e^{i\theta}\tanh r)^k(e^{-i\theta} \sinh r\cosh r)^n }{2^{k+n}(\cosh r)^{(p+1/2)}}\nonumber\\
    &\times\sqrt{\binom{2n}{n}\binom{2k}{k}\binom{p}{2n}\binom{p+2(k-n)}{2k}}\nonumber\\
    &\times\,\ket{p+2(k-n)}.\label{eq:sq1terms}
\end{align}

\subsection{Computing the baselines}\label{app:baselines}
We introduced four novel experiments that optimize the tradeoff between fidelity and success probability.
As the latter was not considered by the baseline experiments, we had to recompute them.
To ensure a meaningful comparison, we optimized crystal sorting in the path identity experiments.
For the heralded Bell, we had to optimize the experiment's parameters for different fidelities.
This section shows the challenges posed by each experiment and how we obtained the baseline curves shown in Fig.~\ref{fig:summary}.

\subsubsection{4-qubit W state}\label{app:bl_w}
\begin{figure}[h]
    \centering
    \includegraphics[width=.99\linewidth]{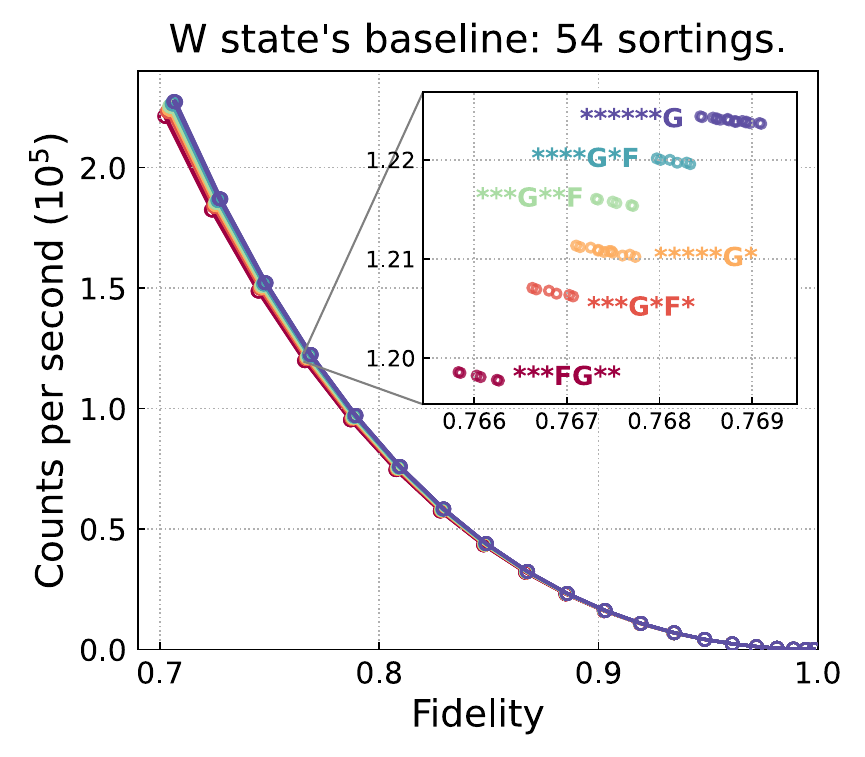}
    \caption{\textbf{54 canonical sortings for the baseline design.} When generating the W state, the sources' ordering in the baseline design leads to different tradeoffs between fidelity and success probability. The positions of sources F and G are a good proxy for performance.
    }
    \label{fig:wstatebl}
\end{figure}

For the W state, the baseline we surpassed was introduced as the following path identity experiment~\cite{graphs3}. % ((0, 7), (0, 5), (0, 3), (2, 4), (2, 6), (1, 2), (4, 6))
\begin{center}
\originalwstate
\end{center}
The squeezing parameter is the same for all 7 sources. From 5040 possible permutations, there are only 54 canonical sortings.
These are obtained from the partial commutation relations between operators and the permutation symmetry of the optical paths.
While the variation between sequences is smaller than in the N00N examples, we see that the 54 designs merge into 6 groups based on performance, which can be predicted from the position of the operators $F$ and $G$ (see Fig.~\ref{fig:wstatebl}).

\subsubsection{Heralded Bell state}\label{app:bl_bell}
The heralded Bell state we surpassed was not presented as an undirected graph, but as the following setup~\cite{sliwa2003conditional}.
\begin{center}
\HeraldedBellCircuitFour
\end{center}
\begin{center}
\HeraldedBellCircuitEight
\end{center}

Equivalently, each optical path can be split into two polarization modes, making the polarizing beam splitters redundant.
Additionally, each of the beam splitters (BS1 and BS2) turns into 2 beam splitters, one per mode, and the half-wave plate becomes a beam splitter between the two modes within the same optical path.

Notice that the two undetected modes in the top (bottom) correspond to a single optical path, like in the heralded experiment in Fig.~\ref{fig:summary}b.
The original photon source was a type-II SPDC~\cite{kok2000postselected}, modeled as
\begin{align}
    S_2(r) =& \exp \left(r (h_1 v_2  - h_1^\dagger v_2^\dagger) - r(h_2 v_1  - h_2^\dagger v_1^\dagger) \right) \label{eq:type2spdc}\\
    =& \exp \left(r (h_1 v_2  - h_1^\dagger v_2^\dagger)\right) \exp \left(- r(h_2 v_1  - h_2^\dagger v_1^\dagger) \right), \nonumber
\end{align}
following the notation of Eq.~\ref{eq:ghz4approx}.
As depicted in the design, this is equivalent of having two sources acting on the same paths with a phase difference of $\pi$.
Thus, the two sources share a common pump $r\in\mathbb R$. Additionally, the beamsplitters (BS1 and BS2) were characterized by a common transmission factor $t$.
To obtain the baseline data shown in Fig.~\ref{fig:summary}, we optimized both parameters for different fidelity values $f_0$ (see Eq.~\ref{eq:loss_function}).
In contrast, the crystals' pumps of the other 2 baselines had a predefined ratio, and we just rescaled them (see Fig.~\ref{fig:ghz4circuits}c and Fig.~\ref{fig:noon4bl}).

\subsubsection{2-mode N00N states}\label{app:bl_noon}
\begin{figure}[!h]
    \centering
    \includegraphics[width=.99\linewidth]{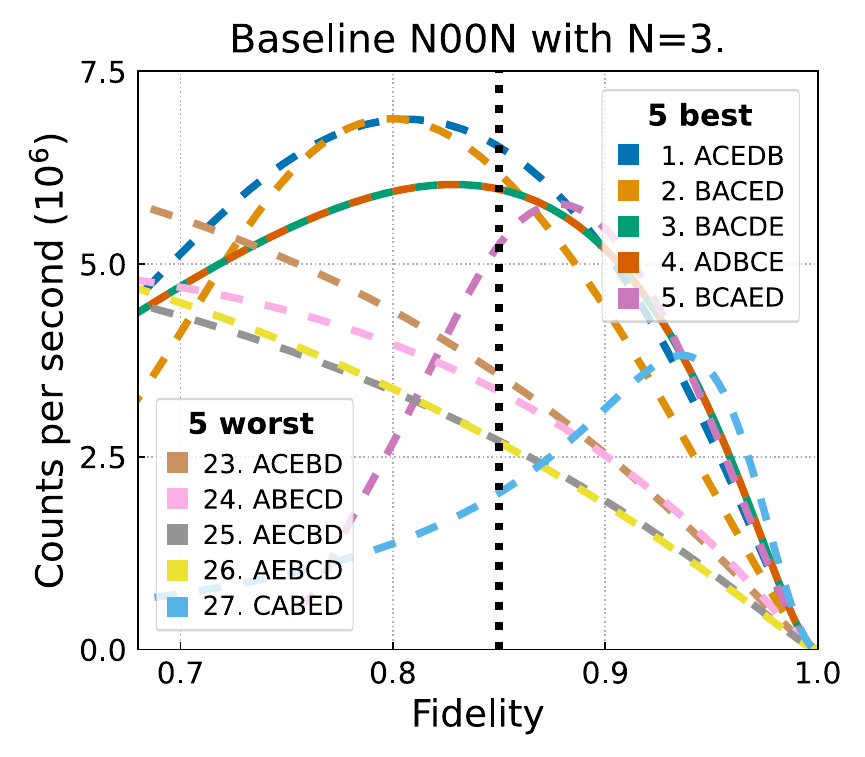}
    \caption{\textbf{Best \& worst sortings for the 3-photon N00N.} We ranked the operators' sortings based on their successful detection counts at fidelity $85\%$. Yet, no sequence is always superior for all fidelity values.}
    \label{fig:noon3bl}
\end{figure}

The N00N state experiments use two main modes and one ancillary mode.
Therefore, without repetition, we have at most 6 sources with 180 unique sequences, 89 if we consider the equivalence between the first 2 modes.
\begin{center}
\sortednoon
\end{center}
The baselines were even smaller~\cite{ruiz2023digital}. For N=3, the baseline used only 5 sources, introduced as the graph:
\begin{center}
\BLnoonThreeGraph
\end{center}
The square node shows the ancillary mode, and the labels the ratio between crystal pumps.
Following the partial commutation rules, the operators can be sorted in 54 distinct ways.
However, as the two main modes can be switched, the canonical sortings are only 27.
While they converge to fidelity 1, the tradeoff curves diverge for lower fidelities.
In Fig.~\ref{fig:noon3bl}, we show the ``best'' and ``worst'' orderings based on the success probability at fidelity 85\%.
Yet, the worst classified reached the highest probability at high fidelity ($>$95\%).

\begin{figure}[]
    \centering
    \includegraphics[width=.99\linewidth]{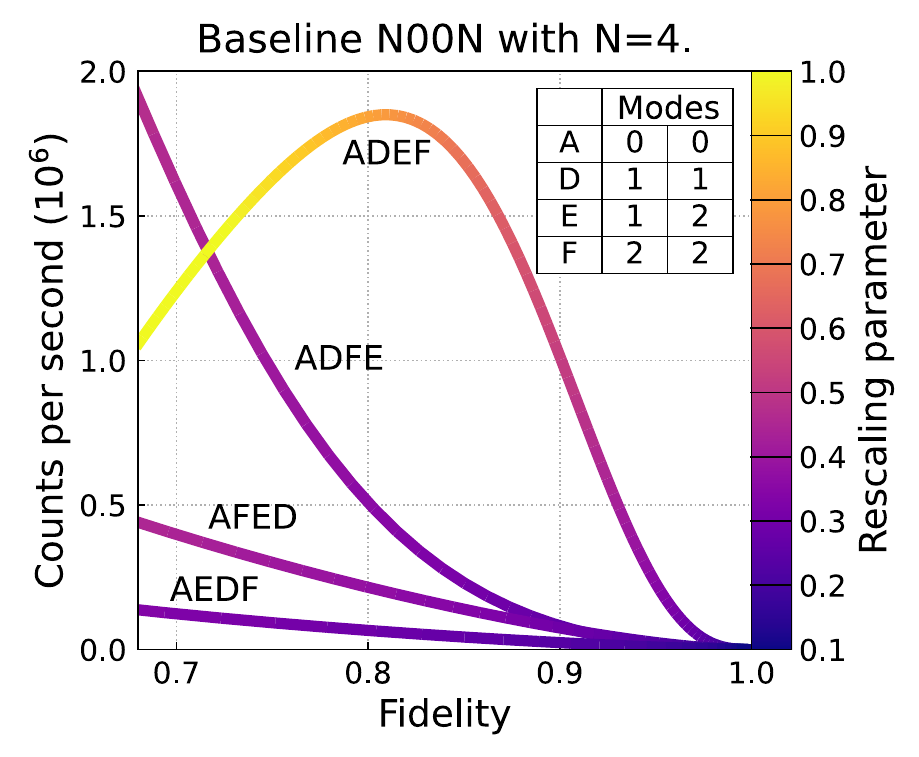}
    \caption{\textbf{All possible sortings for the 4-photon N00N.}
    Keeping the original ratio between sources, we show how the tradeoff changes with the crystal pump.
    Among 4 possible sortings, one stands as the best design for fidelity $>73\%$.}
    \label{fig:noon4bl}
\end{figure}

For N=4, the original baseline used 5 sources and 2 ancillary modes (see the following diagram) 
\begin{center}
\BLnoonFourGraph
\end{center}
Here, we realized that there is no need for a second ancillary mode.
As photons are always created in pairs and we know that the main modes receive exactly 2 photons, a single path with a threshold photodetector suffices.
\begin{center}
\BLnoonFourModifiedGraph
\end{center}
To have a fair comparison, we simplified the baseline into a more efficient graph, with only 4 sources and 1 ancilla. The remaining sources can be sorted in four ways. As shown in Fig.~\ref{fig:noon4bl}, one sorting surpasses the others for all fidelities higher than 73\%.

\newpage

\end{document}